\documentstyle[twoside]{article}

\catcode`\@=11
\long\def\@makefntext#1{
\protect\noindent \hbox to 3.2pt {\hskip-.9pt  
$^{{\eightrm\@thefnmark}}$\hfil}#1\hfill}               %CAN BE USED 

\def\@makefnmark{\hbox to 0pt{$^{\@thefnmark}$\hss}}    %ORIGINAL 
        
\def\ps@myheadings{\let\@mkboth\@gobbletwo
\def\@oddhead{\hbox{}
\rightmark\hfil\eightrm\thepage}   
\def\@oddfoot{}\def\@evenhead{\eightrm\thepage\hfil
\leftmark\hbox{}}\def\@evenfoot{}
\def\sectionmark##1{}\def\subsectionmark##1{}}

\oddsidemargin=\evensidemargin
\newcounter{sectionc}\newcounter{subsectionc}\newcounter{subsubsectionc}
\renewcommand{\section}[1] {\vspace{12pt}\addtocounter{sectionc}{1} 
\setcounter{subsectionc}{0}\setcounter{subsubsectionc}{0}\noindent 
        {\tenbf\thesectionc. #1}\par\vspace{5pt}}
\renewcommand{\subsection}[1] {\vspace{12pt}\addtocounter{subsectionc}{1} 
        \setcounter{subsubsectionc}{0}\noindent 
        {\bf\thesectionc.\thesubsectionc. {\kern1pt \bfit #1}}\par\vspace{5pt}}
\renewcommand{\subsubsection}[1] {\vspace{12pt}\addtocounter{subsubsectionc}{1}
        \noindent{\tenrm\thesectionc.\thesubsectionc.\thesubsubsectionc.
        {\kern1pt \tenit #1}}\par\vspace{5pt}}
\newcommand{\nonumsection}[1] {\vspace{12pt}\noindent{\tenbf #1}
        \par\vspace{5pt}}

\newcounter{appendixc}
\newcounter{subappendixc}[appendixc]
\newcounter{subsubappendixc}[subappendixc]
\renewcommand{\thesubappendixc}{\Alph{appendixc}.\arabic{subappendixc}}
\renewcommand{\thesubsubappendixc}
        {\Alph{appendixc}.\arabic{subappendixc}.\arabic{subsubappendixc}}

\renewcommand{\appendix}[1] {\vspace{12pt}
        \refstepcounter{appendixc}
        \setcounter{figure}{0}
        \setcounter{table}{0}
        \setcounter{lemma}{0}
        \setcounter{theorem}{0}
        \setcounter{corollary}{0}
        \setcounter{definition}{0}
        \setcounter{equation}{0}
        \renewcommand{\thefigure}{\Alph{appendixc}.\arabic{figure}}
        \renewcommand{\thetable}{\Alph{appendixc}.\arabic{table}}
        \renewcommand{\theappendixc}{\Alph{appendixc}}
        \renewcommand{\thelemma}{\Alph{appendixc}.\arabic{lemma}}
        \renewcommand{\thetheorem}{\Alph{appendixc}.\arabic{theorem}}
        \renewcommand{\thedefinition}{\Alph{appendixc}.\arabic{definition}}
        \renewcommand{\thecorollary}{\Alph{appendixc}.\arabic{corollary}}
        \renewcommand{\theequation}{\Alph{appendixc}.\arabic{equation}}
        \noindent{\tenbf Appendix \theappendixc #1}\par\vspace{5pt}}
\newcommand{\subappendix}[1] {\vspace{12pt}
        \refstepcounter{subappendixc}
        \noindent{\bf Appendix \thesubappendixc. {\kern1pt \bfit #1}}
        \par\vspace{5pt}}
\newcommand{\subsubappendix}[1] {\vspace{12pt}
        \refstepcounter{subsubappendixc}
        \noindent{\rm Appendix \thesubsubappendixc. {\kern1pt \tenit #1}}
        \par\vspace{5pt}}

\newcommand{\textlineskip}{\baselineskip=13pt}
\newcommand{\smalllineskip}{\baselineskip=10pt}

\def\eightcirc{
\begin{picture}(0,0)
\put(4.4,1.8){\circle{6.5}}
\end{picture}}
\def\eightcopyright{\eightcirc\kern2.7pt\hbox{\eightrm c}} 

\newcommand{\copyrightheading}[1]
        {\vspace*{-2.5cm}\smalllineskip{\flushleft
        {\footnotesize International Journal of Modern Physics C, #1}\\
        {\footnotesize $\eightcopyright$\, World Scientific Publishing
         Company}\\
         }}

\newcommand{\publisher}[2]{{\begin{center}\footnotesize\smalllineskip 
        Received #1\\
        Revised #2
        \end{center}
        }}

\def\abstracts#1#2#3{{
        \centering{\begin{minipage}{4.5in}\baselineskip=10pt\footnotesize
        \parindent=0pt #1\par 
        \parindent=15pt #2\par
        \parindent=15pt #3
        \end{minipage}}\par}} 

\def\keywords#1{{
        \centering{\begin{minipage}{4.5in}\baselineskip=10pt\footnotesize
        {\footnotesize\it Keywords}\/: #1
        \end{minipage}}\par}}

\newcommand{\bibit}{\nineit}
\newcommand{\bibbf}{\ninebf}
\renewenvironment{thebibliography}[1]
        {\frenchspacing
         \ninerm\baselineskip=11pt
         \begin{list}{\arabic{enumi}.}
        {\usecounter{enumi}\setlength{\parsep}{0pt}     
         \setlength{\leftmargin 12.7pt}{\rightmargin 0pt} %FOR 1--9 ITEMS
         \setlength{\itemsep}{0pt} \settowidth
        {\labelwidth}{#1.}\sloppy}}{\end{list}}

\newcounter{itemlistc}
\newcounter{romanlistc}
\newcounter{alphlistc}
\newcounter{arabiclistc}

\newcommand{\fcaption}[1]{
        \refstepcounter{figure}
        \setbox\@tempboxa = \hbox{\footnotesize Fig.~\thefigure. #1}
        \ifdim \wd\@tempboxa > 5in
           {\begin{center}
        \parbox{5in}{\footnotesize\smalllineskip Fig.~\thefigure. #1}
            \end{center}}
        \else
             {\begin{center}
             {\footnotesize Fig.~\thefigure. #1}
              \end{center}}
        \fi}

\newcommand{\tcaption}[1]{
        \refstepcounter{table}
        \setbox\@tempboxa = \hbox{\footnotesize Table~\thetable. #1}
        \ifdim \wd\@tempboxa > 5in
           {\begin{center}
        \parbox{5in}{\footnotesize\smalllineskip Table~\thetable. #1}
            \end{center}}
        \else
             {\begin{center}
             {\footnotesize Table~\thetable. #1}
              \end{center}}
        \fi}

\def\@citex[#1]#2{\if@filesw\immediate\write\@auxout
        {\string\citation{#2}}\fi
\def\@citea{}\@cite{\@for\@citeb:=#2\do
        {\@citea\def\@citea{,}\@ifundefined
        {b@\@citeb}{{\bf ?}\@warning
        {Citation `\@citeb' on page \thepage \space undefined}}
        {\csname b@\@citeb\endcsname}}}{#1}}

\newif\if@cghi
\def\cite{\@cghitrue\@ifnextchar [{\@tempswatrue
        \@citex}{\@tempswafalse\@citex[]}}
\def\citelow{\@cghifalse\@ifnextchar [{\@tempswatrue
        \@citex}{\@tempswafalse\@citex[]}}
\def\@cite#1#2{{$\null^{#1}$\if@tempswa\typeout
        {IJCGA warning: optional citation argument 
        ignored: `#2'} \fi}}

\def\pmb#1{\setbox0=\hbox{#1}
        \kern-.025em\copy0\kern-\wd0
        \kern.05em\copy0\kern-\wd0
        \kern-.025em\raise.0433em\box0}

\def\fnt#1#2{\footnotetext{\kern-.3em
        {$^{\mbox{\scriptsize #1}}$}{#2}}}

\def\fpage#1{\begingroup
\voffset=.3in
\thispagestyle{empty}\begin{table}[b]\centerline{\footnotesize #1}
        \end{table}\endgroup}

\def\runninghead#1#2{\pagestyle{myheadings}
\markboth{{\protect\footnotesize\it{\quad #1}}\hfill}
{\hfill{\protect\footnotesize\it{#2\quad}}}}
\font\tenrm=cmr10
\font\tenit=cmti10 
\font\tenbf=cmbx10
\font\bfit=cmbxti10 at 10pt
\font\ninerm=cmr9
\font\nineit=cmti9
\font\ninebf=cmbx9
\font\eightrm=cmr8

\def\qed{\hbox{${\vcenter{\vbox{                        %HOLLOW SQUARE
   \hrule height 0.4pt\hbox{\vrule width 0.4pt height 6pt
   \kern5pt\vrule width 0.4pt}\hrule height 0.4pt}}}$}}

\def\bsc{{\sc a\kern-6.4pt\sc a\kern-6.4pt\sc a}}       %LATEX LOGO
\def\bflatex{\bf L\kern-.30em\raise.3ex\hbox{\bsc}\kern-.14em 
T\kern-.1667em\lower.7ex\hbox{E}\kern-.125em X} 

\begin{document}

\runninghead{Individual adaption in a path-based simulation $\ldots$}
{Individual adaption in a path-based simulation $\ldots$}

\normalsize\textlineskip
\thispagestyle{empty}
\setcounter{page}{1}

\copyrightheading{Vol. 0, No. 0 (1993) 000--000}

\vspace*{0.88truein}

\fpage{1}
\centerline{\bf INDIVIDUAL ADAPTION IN A PATH-BASED SIMULATION OF}
\vspace*{0.035truein}
\centerline{\bf THE FREEWAY NETWORK OF NORTHRHINE-WESTFALIA}
\vspace*{0.37truein}
\centerline{\footnotesize KAI NAGEL}
\vspace*{0.015truein}
\centerline{\footnotesize\it Los Alamos National Laboratory, Technology and
  Safety Assessment Division/Simulation}
\baselineskip=10pt
\centerline{\footnotesize\it Applications, Mail Stop M997, Los Alamos NM
  87545, U.S.A., kai@lanl.gov}
\centerline{\footnotesize\it and}
\centerline{\footnotesize\it Santa Fe Institute, 1399 Hyde Park Road, Santa
  Fe NM 87501, U.S.A., kai@santafe.edu}
\centerline{\footnotesize\it }

\vspace*{0.225truein}
\publisher{(received date)}{(revised date)}

\vspace*{0.21truein}

\abstracts{ Traffic simulations are made more realistic by giving
individual drivers intentions, i.e.\ an idea of where they want to go.  One
possible implementation of this idea is to give each driver an exact
pre-computed path, that is, a sequence of roads this driver wants to
follow.  This paper shows, in a realistic road network, how repeated
simulations can be used so that drivers can explore different paths, and
how macroscopic quantities such as locations of jams or network throughput
change as a result of this.}{}{}

\vspace*{10pt}

%\textlineskip                  %) USE THIS MEASUREMENT WHEN THERE IS
%\vspace*{12pt}                 %) NO SECTION HEADING

\vspace*{1pt}\textlineskip      %) USE THIS MEASUREMENT WHEN THERE IS
\section{Introduction}          %) A SECTION HEADING
\vspace*{-0.5pt}

\noindent
It is by now clear that large-scale microsimulations of transportation
systems, with simulation speeds of 1 millions or more vehicles in real
time, are
possible.\cite{Nagel:Schleicher,Rickert:diplom,Rickert:Wagner,Rickert:Wagner:Gawron,PARAMICS:Supercomp}
It is less clear how to ``drive'' these simulations, i.e.\ according
to which rules the individual vehicles know where they are headed.

Random turning choices at intersections, as they would probably be favored
by the Statistical Physics community, do not work well: They are already
unable to represent a simple situation where the traffic in, say,
North-South direction is more important than traffic in the other
directions.

Traffic science traditionally uses turning percentages (see, e.g.,
Ref.~\cite{TRAF}), i.e.\ a table for each direction of each intersection
which says which fraction of vehicles would go left, straight, right, etc.
Apart from the problem of how to collect all the necessary data from the
real world, this is only useful for representing the status quo, but
useless if one wants to study changes in the transportation system, because
the turning percentages change immediately.

The only way out seems to give individual drivers intentions, i.e.\ an idea
of where they want to go.  One possible implementation of this idea is to
give each driver an exact pre-computed ``plan'', i.e.\ a sequence of roads
this driver wants to follow.\cite{TRANSIMS,Rickert:Wagner} See, e.g.,
Refs.~\cite{VanAerde:INTEGRATION2:uguide,VanAerde:96:INTEGRATION} for an
alternative method.

Pre-computed trip plans do not allow an adjustment during the trip.
To make up for this, the simulation can be run several times
(periods/days), and simulated drivers can make different choices each
day, until they settle down on a choice which is favorable for them.
This seems to be a reasonable approach for recurrent (e.g.\ rush-hour)
congestion.\cite{Hall:nonrecurrent,Emmerink:recurrent,Mahmassani:day:to:day}

This paper describes such a simulation using the freeway network of
the German Land Northrhine-Westfalia (NRW).  There are many travelers
with different origin-destination pairs.  Travelers have route plans
(paths) so that they know on which intersections they have to make
turns in order to reach their destinations.  In the simulation setup
described in this paper, each traveler has a choice between
10~different paths.  Each traveler chooses a path, the microsimulation
is executed according to the plans of each traveler (no re-planning
during the trip), and each traveler remembers the performance of
his/her option.

Each traveler tries each option once. Afterwards, she usually chooses
the option which performed best in the past, except that, with a small
probability $p_{other}$, another option is chosen randomly, in order
to update the information about these other options.

This approach---giving each agent a set of options and let each agent act
on the basis of the performances of these options---is a simplified version
of Holland's classifier systems.\cite{Holland:sfi} See
Ref.~\cite{Arthur:bar:problem} for an application of these ideas in an
economic context; and Refs.~\cite{Nagel:Rasmussen,Kelly} for their use in
much simpler transportation problems.

Section~2 describes the digital road network used for the simulations;
section~3 describes in detail the simulation setup.  Section~4 describes
simulation results for the adaption scheme; section~5 discusses several
variations of the basic simulation to demonstrate the robustness of the
results.  Section~6 shows, as one measure of effectiveness, the number of
vehicles which reach their destination as a function of time.  Maybe
counter-intuitively, after everybody has settled down on a choice of path
convenient for her{\em self}, the overall network throughput is lower than
when everybody just drives the geometrically shortest path.  A discussion
concludes the paper.

\section{Network}
\protect\label{network}

\noindent
The simulations are based on a digital version of the freeway network of
NRW, where some lower level highways (Bundesstra{\ss}en) are added in order
to prevent free ends inside the network.  The code is written for parallel
computers using message passing, in principle for an arbitrary numbers of
computational nodes (CPNs).  In practice, two Sparc10 workstations, coupled
via optical link and using PVM~3.2, were used.  This indicates that
experiments such as the one presented in this paper are already possible
with a still modest amount of hardware, and that the consistent use of
parallel supercomputers will allow systematic analysis of much larger
systems.

The network data which is used comes from Rickert,\cite{Rickert:diplom}
(see also Ref.~\cite{Rickert:Wagner}) as an intermediate step of his input
data preparation.  The original data is a list of nodes and a list of
edges, where the list of nodes contains all ramps, junctions, and
intersections, and edges are the connecting segments.  In a first step,
Rickert deletes all nodes of degree two (e.g.\ ramps).  The resulting
network is then distributed on the two workstations.  The heuristic used
for this simply cuts the network in east-west direction such that the
computational load on both workstations is approximately the same.  For
details see Refs.~\cite{Rickert:diplom,Rickert:Wagner,Nagel:etc:VECPAR}.

Apart from the network and the individual trip plans, the simulation is
kept as simple and straightforward as possible.  This includes
oversimplified ramps (see~\cite{Nagel:Rasmussen}) and single directional
lanes, i.e.\ one lane in each direction.  The point of this paper is to
show the application of simulated individual decision making in a
simplified transportation context; a more realistic large scale, path
following traffic simulation is for example documented in
Ref.~\cite{Rickert:Wagner}.

\section{Specific simulation setup}
\label{setup}

\noindent
A simulation run consists of a {\em simulation initialization\/} and {\em
daily iterations}.  During the {\em simulation initialization}, at each
boundary segment 2000~vehicles are queued up to enter the simulation.
(Boundary segments are all segments which lead through the border of NRW
and which are thus connected to the rest of the network only at one end.)
%%Let us also denote by {\em boundary nodes\/}
%%the nodes which connect the boundary segments to the rest of the
%%network.
%%For the simulation set-up, the boundary segments are replaced by
%%{\em{inflow segments}} of equal length $l_{inflow}=2000$. Then, each
%%site of each inflow segment is occupied by a car.  This situation
%%looks like a mega-jam on each inflow segment.  
Each car randomly chooses a destination, which is one of the other boundary
segments.  The probability to choose a certain destination is proportional
to the fourth power of the Cartesian distance between the origin and the
destination segment: $P(destination) \propto (distance)^4$.  Obviously,
taking the fourth power biases this selection towards long trips.  Still
during simulation initialization, each vehicle gets a list of 10~different
paths to reach its destination.  These lists have been pre-calculated for
all occurring origin-destination pairs, and contain the 10~geometrically
shortest paths which do not use the same node twice.${}^{18-20}$
%(Ref.~\cite{Moll:personal}, see Refs.~\cite{Nagel:thesis,Lawler}).

After this general simulation initialization, the {\em daily iterations\/}
are started.  Each daily iteration consists of a {\em preparation phase\/}
and a {\em traffic microsimulation phase}.

During each daily {\em preparation phase}, each vehicle individually
decides which path to use.  In the first day, each vehicle uses the
shortest path; during the subsequent nine days, each vehicle randomly
selects one of the not yet tested options.  Starting at day 10, it usually
selects, as mentioned in the introduction, the option with the best
remembered performance (i.e.\ with the lowest $t_{arriv}$ as defined
below).  Sometimes, with probability $p_{other}=5\%$, it selects one of the
other options to re-test it.

Now, the {\em traffic microsimulation phase\/} of the daily traffic
dynamics starts.  Vehicles are updated according to the Nagel-Schreckenberg
driving logic,${}^{21,22}$
%\cite{Nagel:PhysComp,Nagel:Schreckenberg}
and they change segments when they are at a junction or an intersection.
%%Thus, they enter the
%%network via the the boundary node, where they make their first route
%%decision according to their plans.  
Each vehicle follows its plan until it reaches its destination
segment, and when it reaches the end of that segment, it notes the
arrival time $t_{arriv}$, i.e.\ the current iteration step of the
simulation, which is used as performance criterion for this specific
path.  

%%  When this is the first time this specific path has been used,
%%  this is used as the first guess; for repeated trials of the same
%%  route, two schemes are used:\begin{itemize}
%%  
%%  \item
%%  {\bf Myopic:} Each traveler remembers only the last instance, i.e.
%%  \[
%%  t_{remembered}(option) = t_{arriv} \ ,
%%  \]
%%  where $t_{remembered}(option)$ is the arrival time this car keeps in
%%  memory for this option, and on which it bases subsequent decisions.
%%  
%%  \item
%%  {\bf Averaging:} An exponentially decaying memory, i.e.
%%  \[
%%  t_{remembered}(option) = const \cdot t_{remembered}(option)
%%  %
%%  + (1-const) \cdot t_{arriv} \ .
%%  \]
%%  A value of $const=0.7$ was used; $const=0$ reduces to the myopic
%%  scheme. 
%%  
%%  \end{itemize}
%%  In practice and on the level of the descriptions below, no differences
%%  between both learning schemes were observed.

After all vehicles have reached their destinations and recorded the above
information, the next day is started, where all vehicles have the same
initial position and the same destination as before, but may choose, in the
daily preparation phase, a different path according to the adaption rules
described above.

\section{Adaption results}
\label{adaption}

\noindent
Fig.~\ref{400.1400} shows an example of a simulation after 400~and
1400~seconds.  One clearly sees how the initially empty network is filled
by vehicles coming in from the boundary nodes.  Each pixel in the plot
corresponds to a small region of 30~sites (225~m).  Gray dots denote that
there is at least one car in the region, gray stars mark slightly
over-critical regions, where at least one car has velocity zero, and black
triangles mark jams: The density here is above $0.4$.

\begin{figure}
\fcaption{\label{400.1400}%
Situation after 400~iterations (top) and after 1400~iterations (bottom).
}
\end{figure}

\begin{figure}
\fcaption{\label{1.15}%
{\em Top:} Situation at the ``first day'' after 6000~iterations
(100~minutes), when trips through the network are chosen with a fourth
order preference for long trips, and when all drivers follow the
geometrically shortest path. --- {\em Bottom:} Situation at ``day 15''
after 6000~iterations (100~minutes), for the same initial conditions
as for the top figure, but where drivers have ``learned''.}
\end{figure}

Fig.~\ref{1.15} demonstrates the result of the learning algorithm.  Both
the top and the bottom graph use exactly the same initial configuration of
cars with their individual destinations.  Both graphs are snapshots of the
situation after 6000~seconds (100~minutes).  The top figure shows the
situation when every driver follows the geometrically shortest path.  The
bottom figure shows the situation on day 15, when drivers act according to
their previous experiences, i.e.\ they usually use the path where they were
fastest in the past.  Note some important differences between the figures
(the geographical names are shown in the figures):\begin{itemize}

\item
Drivers learn to use A43 from Wuppertal through the Ruhrgebiet.  A43
is not contained in any shortest long-distance path.

\item Around K\"oln, after the learning also the freeway west of K\"oln is
  crowded. 
  
\item A1 between Wuppertal and Kamen is much more crowded after the
  adaption, with densities above 0.4 at many places.
  
\item At Kamen, there is now not only a jam for people coming from
  Hannover, but also for people coming from the Ruhrgebiet.

\end{itemize}
Generally speaking, people ``learn'' according to the programmed rules
to equilibrate the jams, i.e.\ fast ways around congested areas
vanish.

\section{Robustness results}
\label{robustness}

\noindent
One of the general questions of a simulation like this is how
independent the results are from the specific set-up.  For that
reason, we tested several variations of the simulation.

Day-to-day variations of the general traffic jam patterns are low
after day 15.  As an example, Fig.~\ref{16.linear} top shows the same
situation as in Fig.~\ref{1.15} bottom one simulated day later.

\begin{figure}
\fcaption{\label{16.linear}%
{\em Top:} Same as Fig.~\protect\ref{1.15} (bottom), except that it is
one ``day'' later. --- {\em Bottom:} Same as Fig.~\protect\ref{1.15}
(bottom), except that the distance distribution is linear.}
\end{figure}

It has been reported from other traffic simulations that it is
important how the individual agents remember past information.  For
example, a driver which only remembers the last instance of a trial of
a route instead of some average performance may induce more
oscillations into the system.\cite{Axhausen:personal}  However, in
the simulation setup as described here, using different memory rules
did not lead to any noticeable difference in the simulations.  The
author's intuition is that the stochasticity of the underlying
microscopic driving logic introduces already enough ``fuzziness'' into
the system to avoid such oscillations.  Traditional route choice
models often bundle multiple drivers from the same origin to the same
destination in one packet and do thus not allow for variability
between these.\cite{Emmerink:recurrent}

Also, using a smaller bias towards longer distance destinations
(Fig.~\ref{16.linear} bottom, after adaption) does not change the
overall traffic jam structure.

Obviously, it will be necessary to replace the arbitrary
origin-destination-pattern of these simulations by more realistic
data.  Yet, some of the network bottlenecks seem generic with respect
to transit traffic through NRW: Heavy traffic and congestion between
Wuppertal and  Kamen are well known, and, as one sees, a
consequence of the missing extension of the freeway A4 beyond Olpe.
This extension has since long been planned; but it leads through
environmentally sensitive areas, and it is thus under discussion if it
will ever be built.  Note that the simulation methodology presented
here can be used to evaluate the utility of such an extension, or what
is needed to replace it by improvements along existing paths.  Or,
which traffic streams have to be reduced in order to manage with the
currently existing infrastructure, and how this can be achieved.

The problems near Krefeld are due to the same bottleneck in
North-East/South-West direction.  It is also known that the K\"olner
Ring presents a bottleneck.

Ref.~\cite{Nagel:thesis} contains more detailed descriptions.

\section{Network throughput}
\label{throughput}

\noindent
As a quantitative measure, the accumulated number of cars which have
reached their destination is counted as a function of time.
Fig.~\ref{time} shows one result, for the fourth order distance
distribution. Interestingly, the network performance {\em de\/}creases
after drivers have learned.  The probable explanation is that the network
becomes crowded in a ``balanced'' way after drivers learn, whereas before,
some parts are overcrowded and some are rather empty.  It is, for example,
reasonable to believe that, in Fig.~\ref{1.15} (top), a path from the south
- westward around K{\"o}ln - Wuppertal - Kamen - Hannover has higher
throughput than in Fig.~\ref{1.15} (bottom).
Refs.~\cite{Kelly,Cohen.addition.road,Catoni.equilibrium.paradox} contain
other examples of unexpected or counter-intuitive behavior of traffic
systems. 

\begin{figure}

\fcaption{\label{time}%
Accumulated number of cars which have reached their destination.
``Days''~1 and 15 are shown.  In day 1, all vehicles drive according
to their geometrically shortest path, whereas in day~15 everybody has
some knowledge of the travel time on different paths and usually
chooses the fastest one. --- Interestingly, the network throughput {\em
de\/}creases during the relaxation, indicating that indeed something
like grid-lock occurs not only in urban traffic, but can also occur in
a freeway network.  }

\end{figure}

\section{Discussion}

\noindent
The simulations of this paper use individual learning and route selection
in a simulated traffic system.  This produces a reasonable distribution of
the traffic streams, given the initial origin-destination-assumptions.
Thus, this method is capable to do the equivalent of the static equilibrium
assignment\cite{Sheffi} also for a dynamic and congested situation.

It cannot be expected that the simple assumptions yield a completely
realistic picture of traffic streams; and for an exact comparison with
reality no data was available.  More realistic simulations are the topic of
current work.\cite{TRANSIMS,FVU-NRW} Nevertheless, it is perhaps
astonishing that already such a simple model leads to a reasonable
distribution of traffic streams.  Moreover, the traffic patterns after
adaption are robust against different statistical distributions for the
origin-destination pairs, different learning rules, and different days.
This supports the expectation that already relatively few information on
realistic trip generation will yield rather realistic results.

\nonumsection{Acknowledgments}

\noindent
I learned using paths for traffic microsimulation from Chris Barrett,
who had it already implemented in the demonstration version of
TRANSIMS in 1992.  Christoph Moll helped with the enumeration of the
10~shortest paths for each origin-destination pair.

\nonumsection{References}

%%\bibliographystyle{unsrt}
%%\bibliography{kai,ref}

\begin{thebibliography}{10}
\def\newblock{\relax}

\bibitem{Nagel:Schleicher}
K.~Nagel and A.~Schleicher,
\newblock {\bibit Parallel Computing\/} {\bibbf 20}, 146 (1994).

\bibitem{Rickert:diplom}
M.~Rickert,
\newblock Master's thesis, University of Cologne, K\"oln, Germany, 1994.

\bibitem{Rickert:Wagner}
M.~Rickert and P.~Wagner,
\newblock {\bibit Int.J.Mod.Phys.C\/} {\bibbf 7}, 133 (1996).

\bibitem{Rickert:Wagner:Gawron}
M.~Rickert, P.~Wagner, and Ch. Gawron,
\newblock in {\bibit Proceedings of the 4th PASA Workshop 1996\/} 
\newblock (in press).

\bibitem{PARAMICS:Supercomp}
G.D.B.~Cameron and C.I.D.~Duncan,
\newblock {\em J. Supercomputing\/} (in press).

\bibitem{TRAF}
\newblock {\bibit TRAF user reference guide}, 
U.S.Department of Transportation, Federal Highway Administration,
\newblock Publication No.\ FHWA-RD-92-060 (1992).

\bibitem{TRANSIMS}
TRANSIMS, TRansportation ANalysis and SIMulation System, Los Alamos National
  Laboratory, Los Alamos, U.S.A. See {http://www-transims.tsasa.lanl.gov}.

\bibitem{VanAerde:INTEGRATION2:uguide}
M.~Van~Aerde et~al.,
\newblock {\bibit INTEGRATION--Release 2 User's Guide}, 1995.

\bibitem{VanAerde:96:INTEGRATION}
M.~Van Aerde, B.~Hellinga, M.~Baker, and H.~Rakha,
\newblock INTEGRATION: An overview of traffic simulation features,
\newblock {\bibit Transportation Research Records\/} (in press).

\bibitem{Hall:nonrecurrent}
R.W.~Hall,
\newblock {\bibit Transportation Research C\/} {\bibbf 36}, 67 (1994).

\bibitem{Emmerink:recurrent}
R.H.M. Emmerink, K.W. Axhausen, P.~Nijkamp, and P.~Rietveld,
\newblock {\bibit Transportation\/} {\bibbf 22}, 21 (1995).

\bibitem{Mahmassani:day:to:day}
H.S.~Mahmassani, G.-L.~Chang, and R.~Herman,
\newblock {\bibit Transportation Science\/} {\bibbf 20}, 258 (1986).

\bibitem{Holland:sfi}
J.H.~Holland,
\newblock in {\bibit Lectures in the sciences of complexity}, ed.\  
D.~Stein (Addison-Wesley Longman, 1989).

\bibitem{Arthur:bar:problem}
W.B.~Arthur,
\newblock {\bibit American Economic Review (Papers and Proceedings)\/}
{\bibbf 84} (1994).

\bibitem{Kelly}
T.~Kelly, {\bibit Physica A\/} (in press).

\bibitem{Nagel:Rasmussen}
K.~Nagel and S.~Rasmussen,
\newblock in {\bibit Artificial Life IV:
  Proceedings of the Fourth International Workshop on the Synthesis and
  Simulation of Living Systems}, ed.\ R.A.~Brooks and P.~Maes 
(MIT Press, Cambridge, MA, 1994), p.~222.

\bibitem{Nagel:etc:VECPAR}
K.~Nagel, M.~Rickert, and C.L.~Barrett,
\newblock Large scale traffic simulations,
\newblock in {\bibit Proceedings of VECPAR'96\/} (in press).

\bibitem{Moll:personal}
C.~Moll,
\newblock personal communication. 

\bibitem{Nagel:thesis}
K.~Nagel,
\newblock {\bibit High-speed microsimulations of traffic flow},
\newblock PhD thesis, University of Cologne, K{\"o}ln, Germany, 1995.

\bibitem{Lawler}
E.L.~Lawler,
\newblock {\bibit Compinatorial optimization: {N}etworks and matroids\/}
\newblock (Holt, Rinehard, and Winston, New York, 1976).

\bibitem{Nagel:PhysComp}
K.~Nagel, 
\newblock in {\bibit Physics Computing~'92}, 
ed.\ R.A.~de~Groot and J.~Nadrchal (World Scientific, 1993), p.~419.

\bibitem{Nagel:Schreckenberg}
K.~Nagel and M.~Schreckenberg,
\newblock {\bibit J. Phys. I France} {\bibbf 2}, 2221 (1992).

\bibitem{Axhausen:personal}
K.~Axhausen,
\newblock personal communication.

\bibitem{Cohen.addition.road}
J.~Cohen and F.~Kelly, {\bibit J.\ Appl.\ Probability\/} {\bibbf 27}, 730
(1990). 

\bibitem{Catoni.equilibrium.paradox}
S.~Catoni and S.~Pallottino, {\bibit Transportation Science\/} {\bibbf 25},
240 (1991).

\bibitem{Sheffi}
Y.~Sheffi,
\newblock {\bibit Urban transportation networks: Equilibrium analysis with
  mathematical programming methods\/} 
\newblock (Prentice-Hall, Englewood Cliffs, NJ, 1985).

\bibitem{FVU-NRW}
Forschungsverbund f{\"u}r Verkehr und Umwelt (FVU) NRW. See
  {http://www.zpr.uni-koeln.de/Forschungsverbund-Verkehr-NRW/}.

\end{thebibliography}

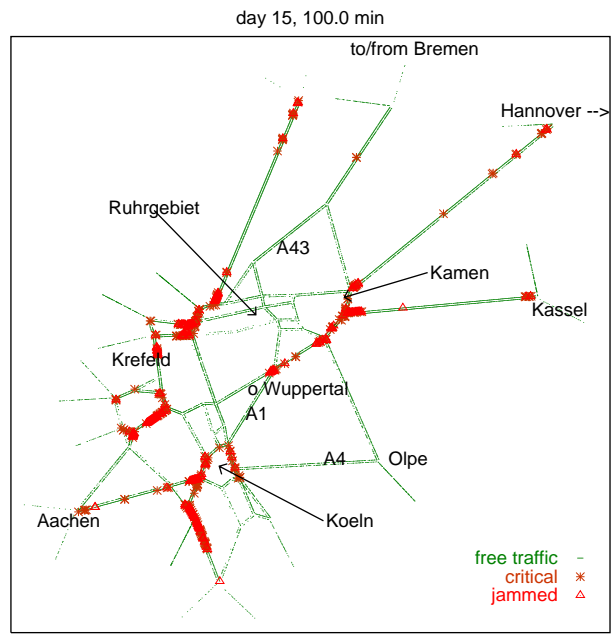
\end{document}